\documentclass{PoS}

\title{Single Top Quark Production at the Tevatron}

\ShortTitle{Single Top Quark Production at the Tevatron} 

\author{\speaker{Yvonne Peters}\thanks{On behalf of the CDF and D0 collaborations.}\\
        University of G\"ottingen and DESY, Germany\\
        E-mail: \email{reinhild.peters@cern.ch}}

\abstract{While the heaviest known elementary particle, the top quark, has been
discovered in 1995 by the CDF and D0 collaborations in $t\bar{t}$
events, it took 14 more years until the observation of single top
quark production. In this article, we discuss recent  studies of single top
quark production by the CDF and D0 collaborations at the Tevatron.
In particular, we present the measurement of single top quark $s$- and
$t$-channel production combined, the first observation of $t$-channel
production, the simultaneous measurement of $s$- and $t$-channel
production cross sections  as
well as the extraction of the CMK matrix element $|V_{tb}|$.}

\FullConference{36th International Conference on High Energy Physics,\\
		July 4-11, 2012\\
		Melbourne, Australia}

\begin{document}

\section{Introduction}
Single top quark production via the electroweak interaction occurs via
the $s$-channel, $t$-channel and $Wt$-channel. The latter has a
negligible cross section at the Tevatron. 
While the discovery of top quarks by the CDF and D0 collaborations
happened  in 1995~\cite{CDF_obs,D0_obs} in $t\bar{t}$ production, it
took additional 14 years to observe single top quark production. Both
the CDF and
D0 collaborations reported single top quark observation in Spring
2009~\cite{cdfsingletop,d0singletop}. For the single top quark
observation, the $s+t$-channel cross
section ($\sigma_{s+t}$) was measured by CDF and D0 using up to $3.2$~fb$^{-1}$ and
$2.3$~fb$^{-1}$ of data, respectively. While the cross section of the
$s+t$-channel is only about a factor of two smaller than $t\bar{t}$
production, its signature is very similar to $W$+jets events, making
single top a challenging process to measure. 

In the following, we present updated measurements of the $s+t$ channel
cross section by CDF and D0, as well as the individual meaurement of
$t$-channel cross section, the simultaneous extraction of $s$- and
$t$-channel cross sections as well as the extraction of the Cabibbo-Kobayashi-Maskawa (CKM) matrix
element $|V_{tb}|$ from single top.

\section{\boldmath Selection and ($s+t$)-channel Measurement}
\begin{figure*}[t]
\centering
\includegraphics[height=50mm]{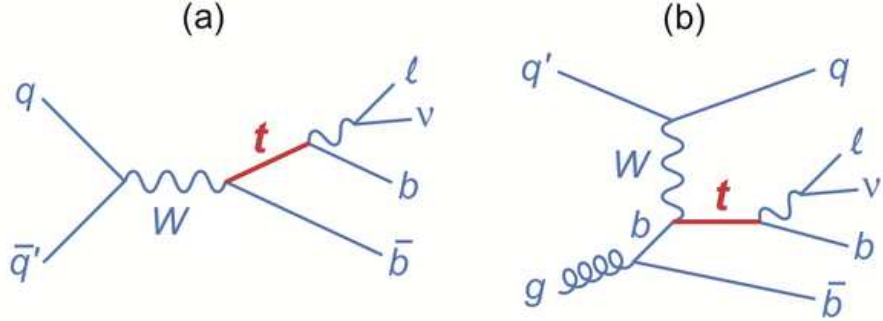}
\caption{Tree-level  Feynman diagrams for $s$-channel (a)
and $t$-channel (b) single top quark production. } \label{feynmansingletop}
\end{figure*}

At the Tevatron $p\bar{p}$ collider with a center of mass energy of
$1.96$~TeV, the cross section of $s$-channel and $t$-channel single
top quark production 
is $1.04$~pb and $2.26$~pb in the standard model (SM),
respectively~\cite{singletoptheory}. Figure~\ref{feynmansingletop}
shows the leading order Feynman diagrams for the $s$-channel (a)
and $t$-channel (b) single top quark production.
The signature of the signal is
very similar to that of $W$ boson production in association with
jets, presenting a challenge for the single top quark cross section  measurements. 
Other background contributions to single top quark measurements result from
$t\bar{t}$ production and instrumental background from jets faking
isolated leptons, and smaller contributions from diboson ($WW$, $WZ$
and $ZZ$) and $Z$+jets production.
The selection of single top quark events focuses on events where the
$W$ boson from the top quark decays into a charged electron or muon
and the corresponding neutrino. Therefore, the requirements are a
charged, isolated lepton (electron or muon) with high transverse
momentum ($p_T$), two, three or four jets, of which the leading jet
has to have high $p_T$, and large missing transverse energy, a characteristic of events with undetected neutrinos. Additionally, cuts on angular
distributions are applied in order to reduce the
instrumental background. Such a cut is for example applied on the angular difference between the transverse lepton direction
and the missing transverse energy at D0. 
 In $t$-channel production, one of the jets
can have high $\eta$ and low $p_T$, which is challenging to model. 

After this pre-selection, the $W$+jets and instrumental backgrounds are estimated using a
data-driven approach. To further increase the purity of the sample,
the application of $b$-jet identification is necessary. A neural-net
based $b$-tagger is used at D0~\cite{btagd0} which is based on
lifetime and displaced track information. At CDF, a secondary vertex
$b$-tagger is used. We require one or two $b$-jets in the event. In
total, up to six channels are considered: The D0 collaboration
considers events with two, three and
four jets, split into events with one or two identified $b$-jets,
while in the recent analysis by the CDF collaboration events with
exactly two or three jets, split into events with one or two
identified $b$-jets are considered.

In order to further enrich the data sample in single top quark events,
multivariate analysis (MVA)  techniques are necessary, that separate
signal from background. 
In the recent single top measurement by CDF, using 7.5~fb$^{-1}$ of
Run~II data, a neural network is trained on single top quark signal. At D0,
using 5.4~fb$^{-1}$ of data, three different multivariate analysis
techniques are applied separately on the data, and these are used as
input to a super-discriminant to increase the sensitivity of the
analysis. In particular, a boosted decision tree (BDT), a bayesian neural
network (BNN) as well as a NEAT (neuro evolution of augmented
topologies) discriminant are applied on the data. For
the reduction of the correlation between the methods, the BNN is
trained using the four-vectors of all final state particles, the
two-vector of the missing transverse energy and a few
additional variables that include lepton charge and $b$-tagging information, while the BDT approach uses a variety of complex
variables as input to the BDT. The super-discriminant is build using a
BNN. 
Depending on the final channel of interest, the training of the MVAs
is performed by training either on $s+t$ channel as signal, by
training on $t$-channel as signal and treating $s$-channel as
background in the training procedure, or by training on $s$-channel as
signal and treating $t$-channel as background during the
training. The CDF collaboration performs the training differently in
the different data samples: for the various jet and $b$-tag bins the
largest expected contribution of $s$- and $t$-channel is estimated and
the training performed on this particular channel in the considered
jet and $b$-tag bins. 

\begin{figure*}[t]
\centering
\includegraphics[width=48mm]{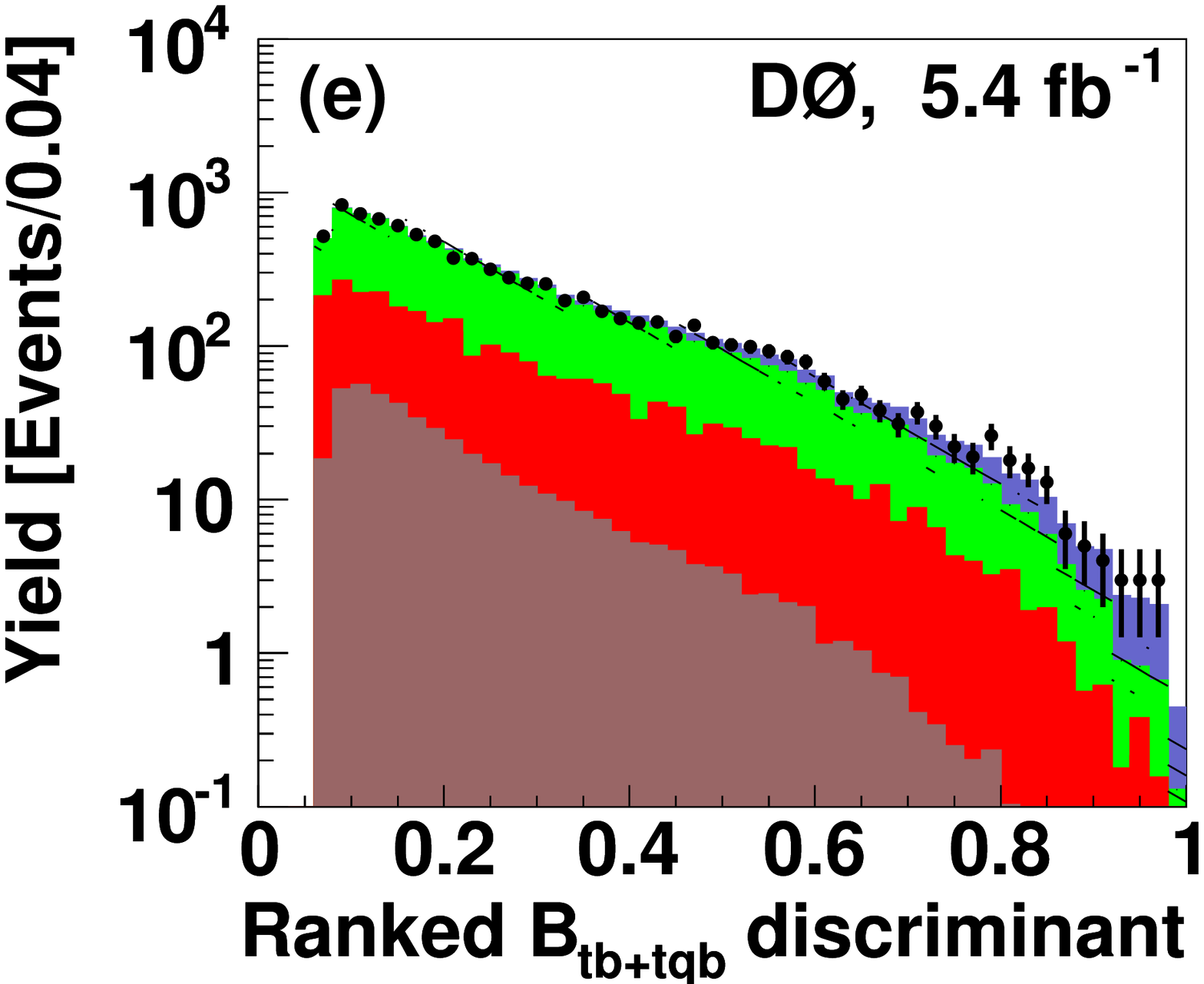}
\includegraphics[width=48mm]{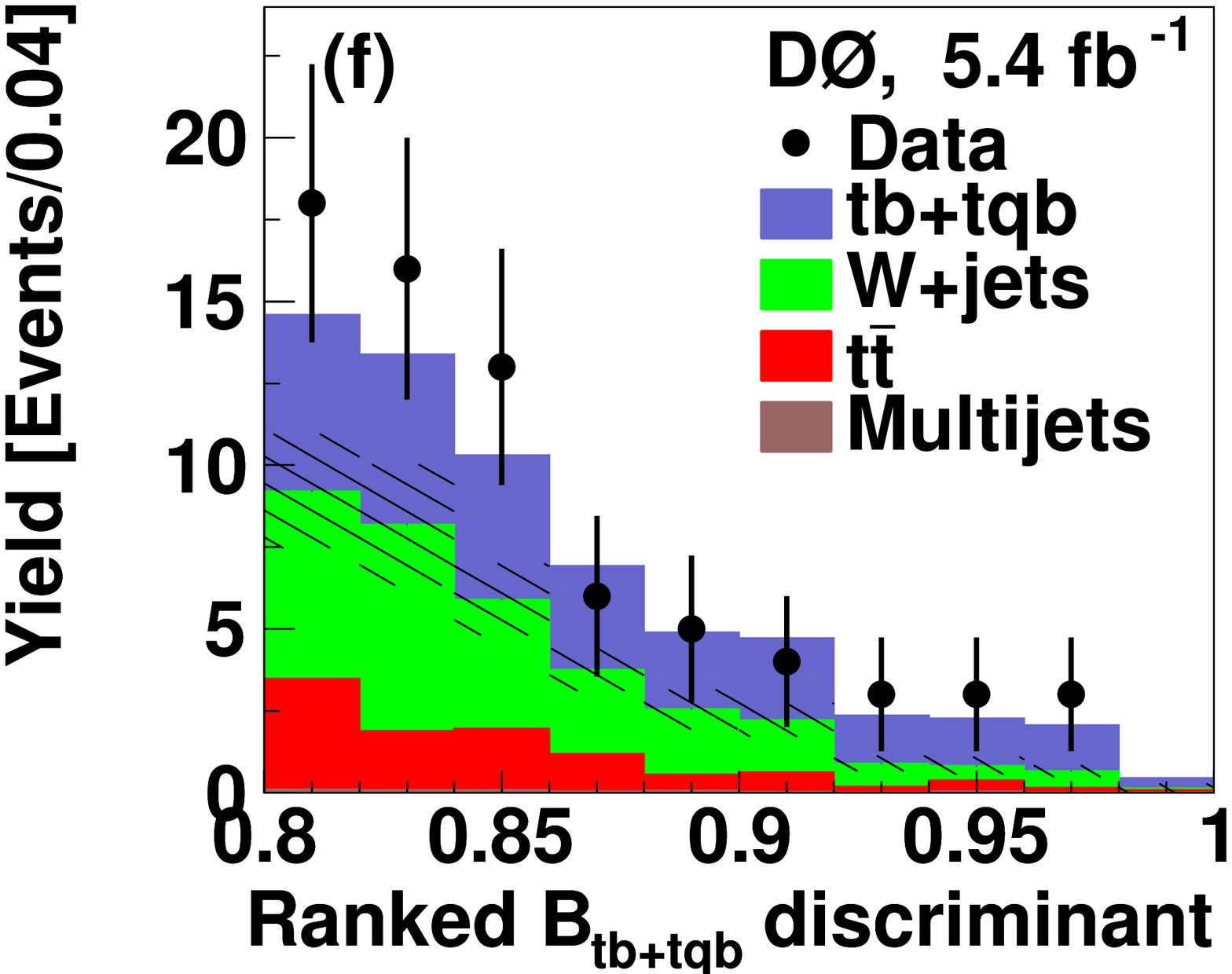}
\includegraphics[width=48mm]{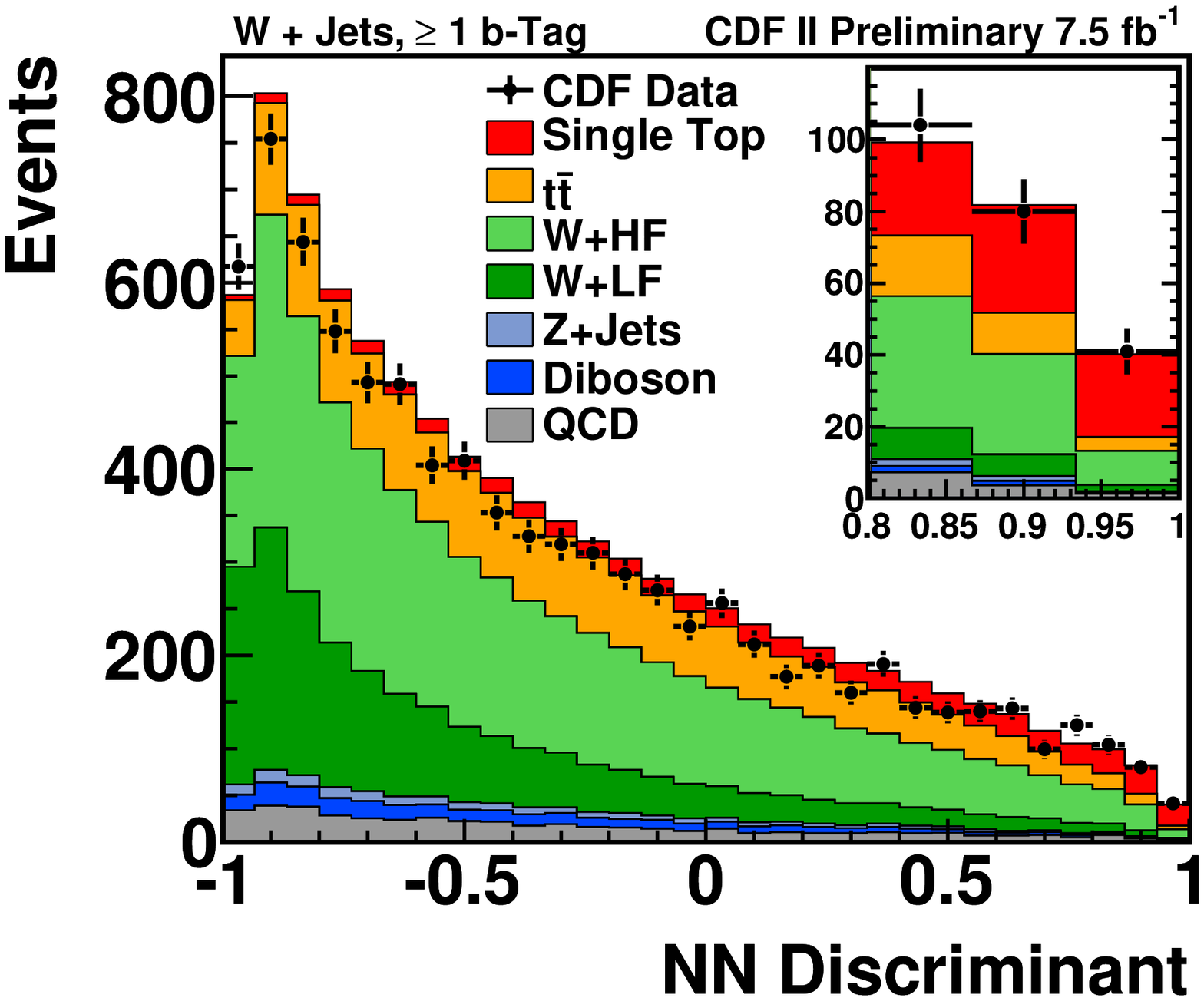}
\caption{Comparison for Data and prediction of the discriminant
  outputs for the combined channel at D0 (left, middle)~\cite{d0newsingletop} and CDF (right)~\cite{cdfnewsingletop}.
The discriminants for D0 are trained on the $s+t$-channel single top
signal, and are ranked according to the signal over background ratio
in the bins. The left plot shows the full distribution, and the middle
plot the zoomed region of high ranked discriminant output. } \label{stsingletop}
\end{figure*}

The final extraction of the cross section from the discriminant is performed using a
Bayesian method, where systematic uncertainties are incorporated as
Gaussian priors and are integrated over. The measured $s+t$-channel cross
section is $\sigma_{s+t}=3.43^{+0.73}_{-0.74}$~pb at D0, where the
training is performed on the $s+t$-channel
as signal, and $\sigma_{s+t}=3.04^{+0.57}_{-0.53}$  at CDF~\cite{d0newsingletop,cdfnewsingletop}. Both
measurements are using a top quark mass of $172.5$~GeV in the Monte
Carlo simulation of the signal. Figure~\ref{stsingletop} shows the
comparison of prediction and data of the 
discriminant outputs for the D0 (left and middle) and CDF (right) measurements.
The main
systematic uncertainties arise from the uncertainty on the integrated
luminosity, from uncertainties on the jet energy scale and resolution,
and on the identification of $b$-jets. Both measurements are in
agreement with each other and with the SM prediction.

%something on Mc modeling? and bkg and such? what generators used for signal?
%check the training for CDF

\section{\boldmath $t$-channel Observation and 2D Measurements}
\begin{figure*}[t]
\centering
\includegraphics[height=68mm]{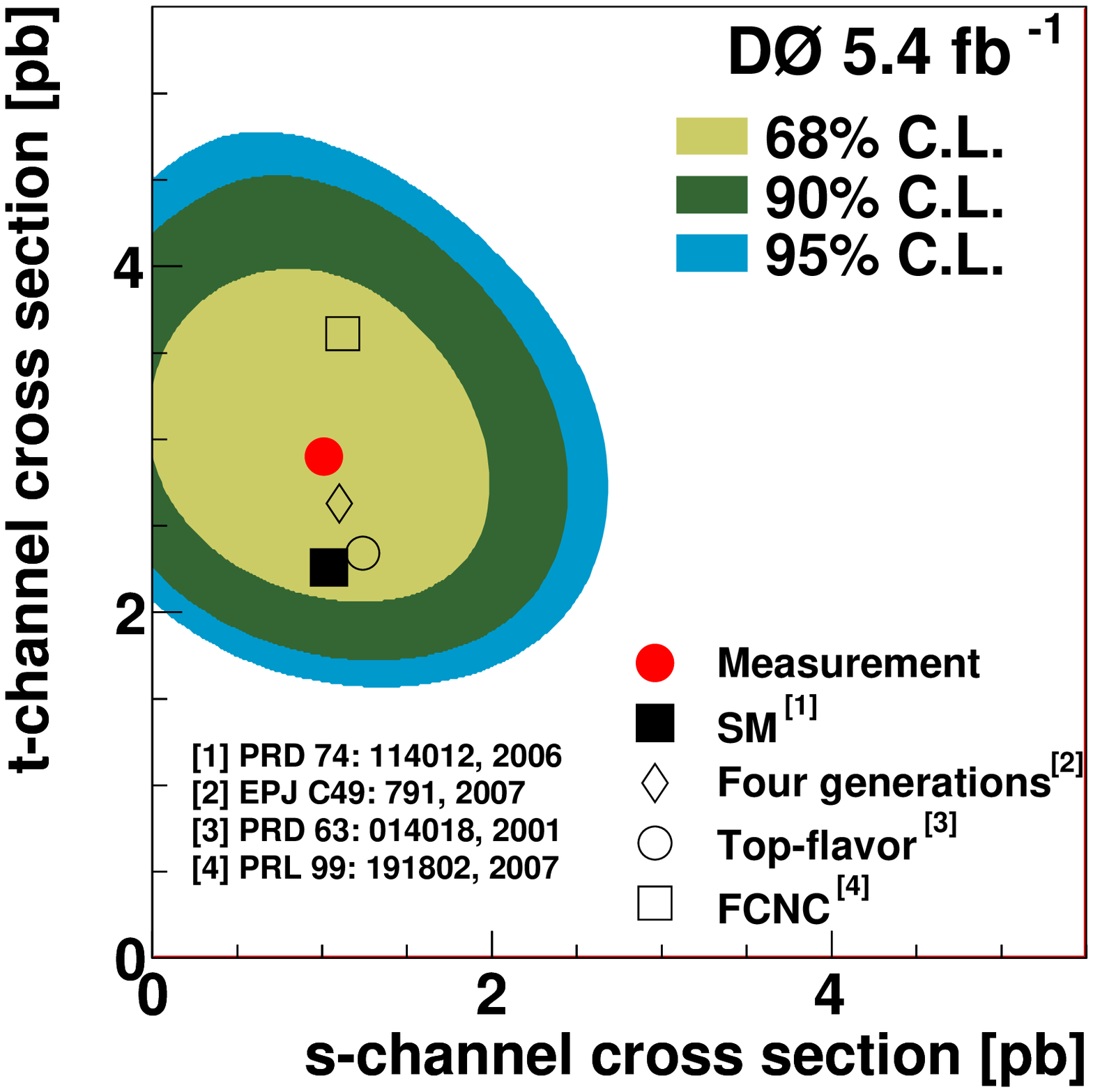}
\includegraphics[height=63mm]{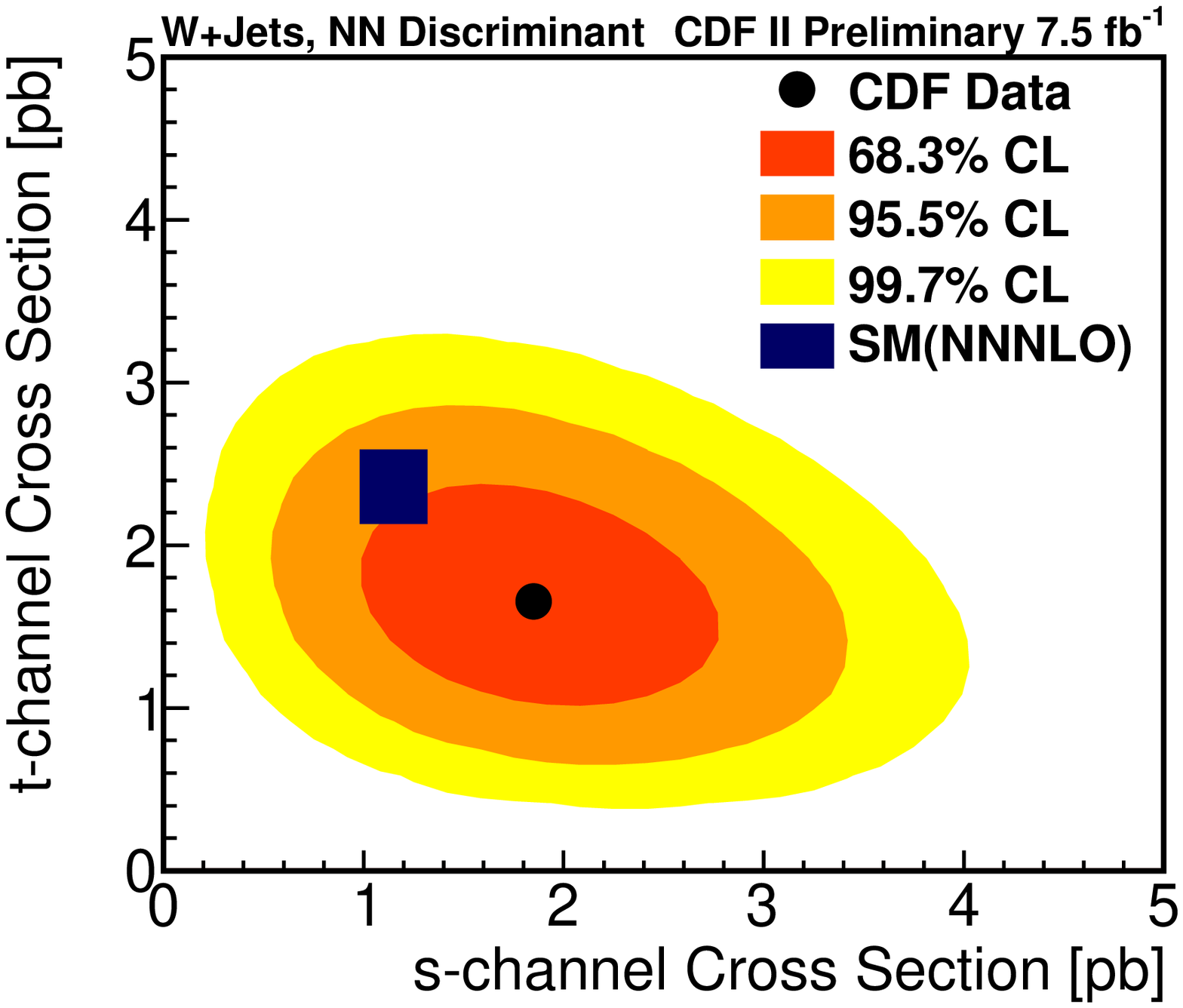}
\caption{Posterior probability density for $s$- versus $t$-channel
  single top quark production cross sections. Contours of equal
  probabilty are shown, plus the measured and predicted cross
  sections, for D0 (left)~\cite{d0tchannelobservation} and CDF
  (right)~\cite{cdfnewsingletop}. Note the difference in choice of
  contours between the two plots.} \label{2dsingletop}
\end{figure*}

Signatures of models for physics beyond the SM could look similar to
$s$-channel or $t$-channel processes. Depending on the choice of
model, either $s$- or $t$-channel optimized selections would be more
sensitive to new physics. Therefore, the simultaneous measurement of
$s$- versus $t$-channel production cross sections is a good way to
probe for new physics. Both, CDF and D0 collaborations, performed the
2D cross section measurement, extracting $s$- and $t$-channel
simultaneously, without assuming the SM ratio between both channels as
is done for the $s+t$-channel extraction. While at D0, the training was performed
with $t$-channel as signal, CDF performs the training on $s$- and
$t$-channel depending on the number of jets and $b$-tags in the
event. Using $5.4$~fb$^{-1}$ of data, D0 measures $\sigma_s=0.98 \pm
0.63$~pb and $\sigma_{t}=2.90 \pm 0.59$~pb~\cite{d0newsingletop}.
The CDF collaboration extracts $\sigma_s=1.81^{+0.63}_{-0.58}$~pb and
$\sigma_t=1.49^{+0.47}_{-0.42}$~pb using 7.5~fb$^{-1}$ of data~\cite{cdfnewsingletop} . Both results are compatible with
the SM prediction.  Due to the different training, CDF is more
sensitive to $s$-channel single top production than D0. In
Fig.~\ref{2dsingletop} the SM cross sections, the measured cross
sections and the contours of equal probability are shown for the $s$- versus $t$-channel
  single top quark production cross section measurement from D0 (left)
  and CDF (right).

By integrating over the $s$-channel, the D0 collaboration also extracts the
$t$-channel cross section from the 2D measurement, resulting in
$\sigma_{t}=2.90 \pm 0.59$~\cite{d0tchannelobservation}.  This result
deviates with more than 5.5 standard deviations from zero, and is the
first observation of the $t$-channel single top cross section. The
result is in good agreement with the SM prediction.

%motivation of different np in both channels
%maybe example model?
% put in 2D plots

\section{\boldmath Measurement of $|V_{tb}|$}
\begin{figure*}[t]
\centering
\includegraphics[height=55mm]{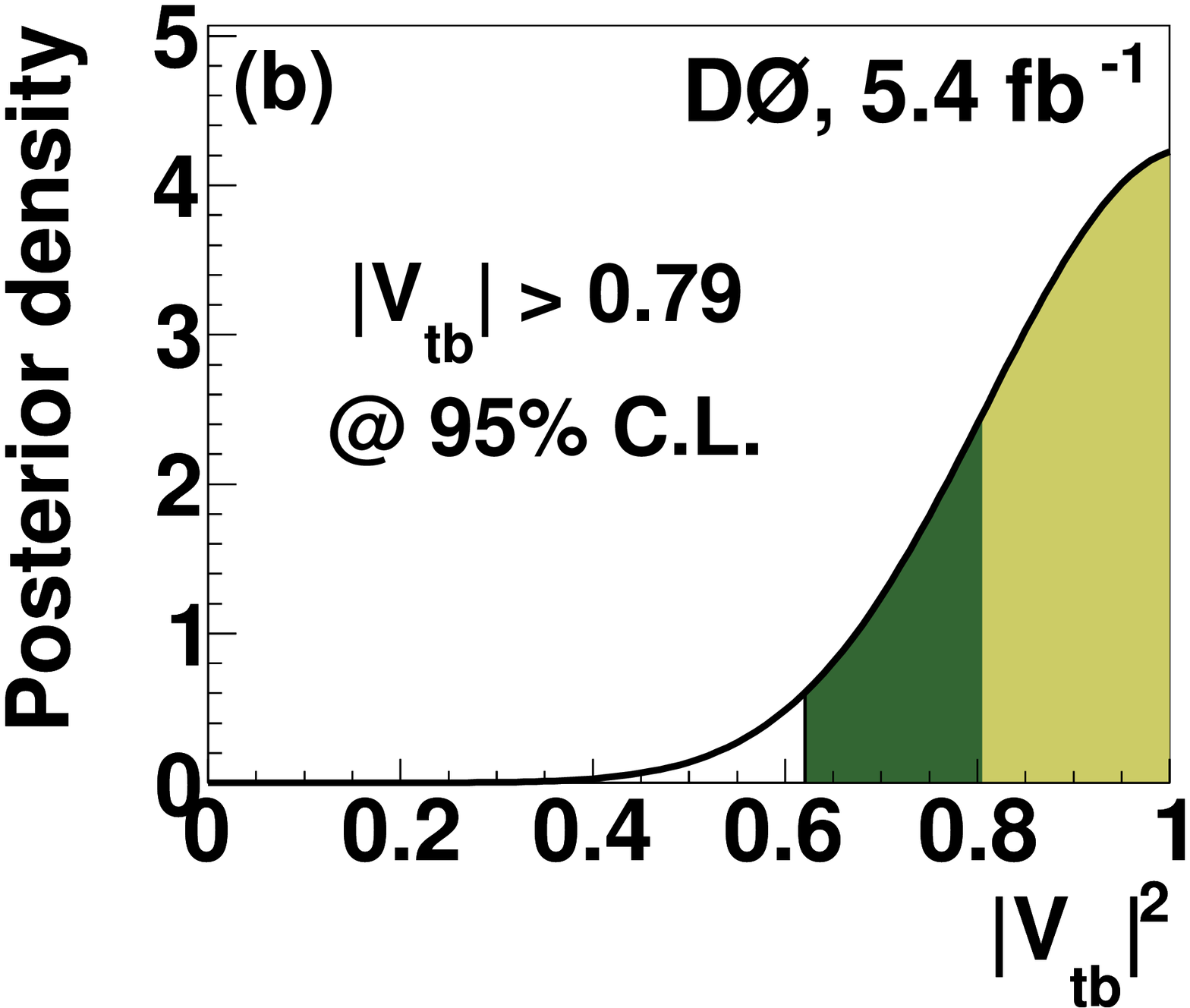}
\includegraphics[height=55mm]{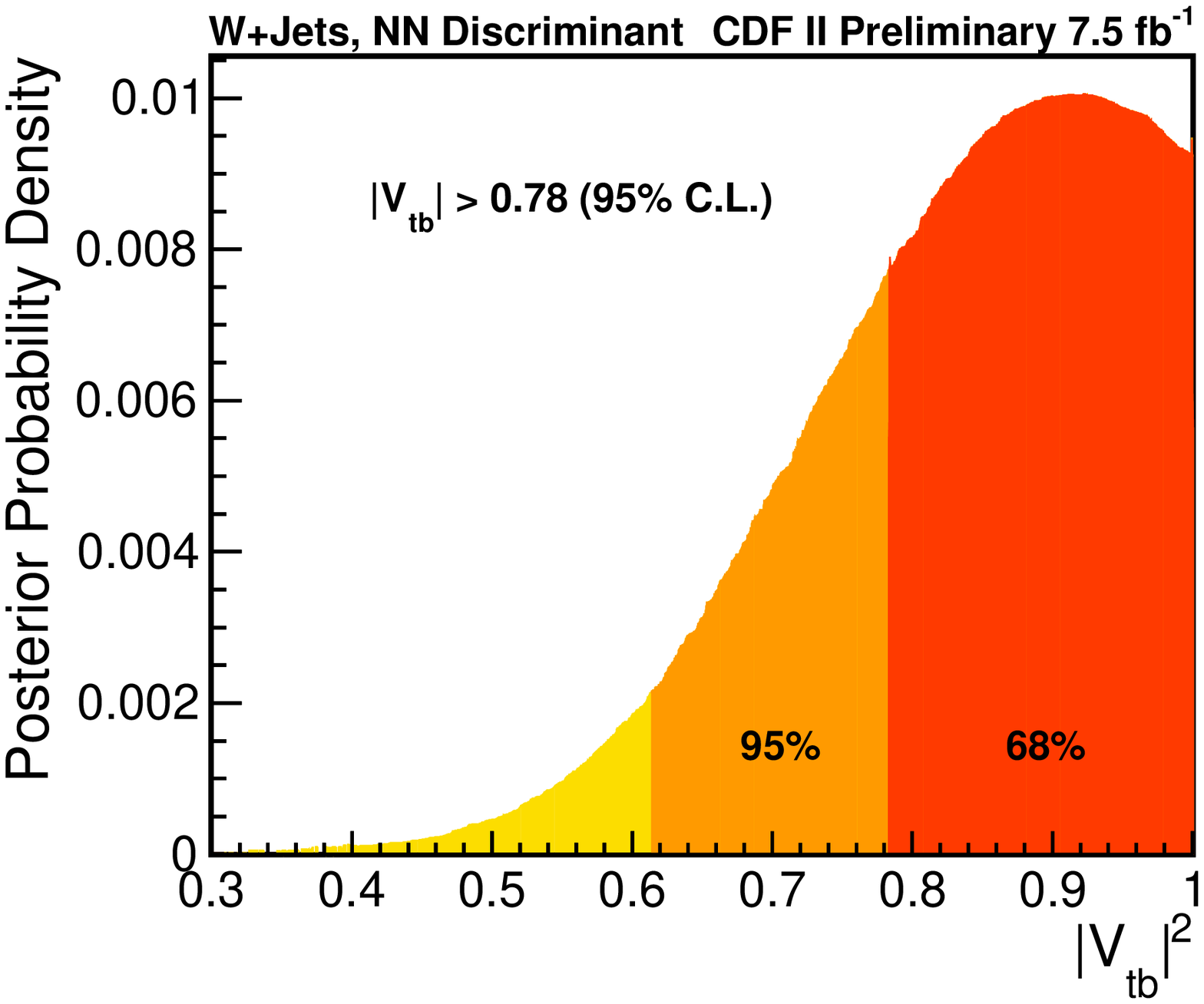}
\caption{The posterior density functions for $|V_{tb}|^2$, for the
  measurement by the D0
  (left)~\cite{d0newsingletop} and CDF (right)~\cite{cdfnewsingletop} collaborations.} \label{vtbsingletop}
\end{figure*}

One interesting application of the single top quark analyses
is the direct extraction of the CKM matrix element $|V_{tb}|$, without
any assumptions about the number of fermion generations. The square of
$|V_{tb}|$ is proportional to the single top cross section
$\sigma_{s+t}$. Using the results for the $s+t$-channel single top
cross section extracted from 5.4~fb$^{-1}$ of D0 and 7.5~fb$^{-1}$ of
CDF data, we extract lower limits of $|V_{tb}|>0.79$ at D0
and $|V_{tb}|>0.78$ at CDF at the 95\% confidence level using a Bayesian
method. In Figure~\ref{vtbsingletop} the posterior curves for
$|V_{tb}|^2$ are shown for the measurement from the D0 (left) and CDF
(right) collaborations. 

While the described extraction of $|V_{tb}|$ is not using any
assumption about the number of quark generations, it does assume
$|V_{ts}|^2, |V_{td}|^2 << |V_{tb}|^2$. An alternative way to extract
$|V_{tb}|$ is by using the result from the indirect determination of
the top quark width~\cite{topwidthpaper,toppropichep}. $|V_{tb}|^2$ is
proportinal to the top width, the latter of which is determined by
combining the measurement of the ratio of branching fractions $R=B(t
\rightarrow Wb)/B(t \rightarrow Wq)$, with $q$ any down-type quark, in
$t\bar{t}$ and the $t$-channel single top cross section. This
$|V_{tb}|^2$ extraction does not assume $|V_{ts}|^2, |V_{td}|^2 <<
|V_{tb}|^2$, and also does not assume the SM ratio between the $s$-
and $t$-channel cross sections. Using the result from the indirect top
width determination, we extract $|V_{tb}|>0.81$.

% Wtb interaction vertex...

\section{Conclusion and Outlook}
The measurement of single top quark production at the Tevatron is
very challenging due to the small production cross section and the
similarity of the single top signal and the main background from  $W$+jets production. Using a variety of
multivariate analysis techniques, both CDF and D0 collaborations
measured single top production cross sections for $t$-channel and
$s+t$-channels. The simultaneous extraction of the  $s$- and $t$-channel
cross section yields a sensitive probe for new physics. Furthermore,
the CKM matrix element $|V_{tb}|$ has been extracted. The full
Tevatron dataset has still to be explored, with the main goal to
get evidence or observation for  the $s$-channel single top quark production.

\section*{Acknowledgements}
I thank my collaborators from CDF and  D0
 for their help in preparing the presentation and this
article. I also thank the staffs at Fermilab and
collaborating institutions, and acknowledge the support from the
Helmholtz association.

\end{document}